\documentclass[10pt]{iopart} %when \ioptwocol
\usepackage{iopams} %IOP specific AMS fonts, symbols

\usepackage{hyperref}
\usepackage[dvipsnames]{xcolor}
\usepackage{cuted} %for 1-column equations (strip)
\usepackage{graphicx,subfig} %for subfigures
\usepackage{wrapfig}
\begin{document}

\title{Total Fluid Pressure Imbalance in the Scrape-Off Layer of Tokamak Plasmas}
\author[1]{R.M. Churchill$^1$, J.M. Canik$^2$, C.S. Chang$^1$, R. Hager$^1$, A.W. Leonard$^3$, R. Maingi$^1$, R. Nazikian$^1$, D.P. Stotler$^1$}

\address{$^1$ Princeton Plasma Physics Laboratory, 100 Stellarator Road, Princeton, NJ 08540, USA}
\address{$^2$ Oak Ridge National Laboratory, PO Box 2008, Oak Ridge, TN 37831, USA}
\address{$^3$ General Atomics, PO Box 85608, San Diego, CA 92186-5608, USA}
%email of corresponding author
\ead{rchurchi@pppl.gov}

%\begin{document}

\begin{abstract}
Simulations using the fully kinetic neoclassical code XGCa were undertaken to explore the impact of kinetic effects on scrape-off layer (SOL) physics in DIII-D H-mode plasmas. XGCa is a total-$f$, gyrokinetic code which self-consistently calculates the axisymmetric electrostatic potential and plasma dynamics, and includes modules for Monte Carlo neutral transport.

Previously presented XGCa results showed several noteworthy features, including large variations of ion density and pressure along field lines in the SOL, experimentally relevant levels of SOL parallel ion flow (Mach number$\sim$0.5), skewed ion distributions near the sheath entrance leading to subsonic flow there, and elevated sheath potentials [R.M. Churchill, \emph{Nucl. Mater. \& Energy}, submitted].

In this paper, we explore in detail the question of pressure balance in the SOL, as it was observed in the simulation that there was a large deviation from a simple total pressure balance (the sum of ion and electron static pressure plus ion inertia). It will be shown that both the contributions from the ion viscosity (driven by ion temperature anisotropy) and neutral source terms can be substantial, and should be retained in the parallel momentum equation in the SOL, but still falls short of accounting for the observed fluid pressure imbalance in the XGCa simulation results.
\end{abstract}

\maketitle
\ioptwocol

\section{Introduction}
Scrape-off layer (SOL) physics in tokamak devices are typically simulated using fluid codes, due to the generally high collisionality in this region. However, research has revealed a number of discrepancies between experiment and leading SOL fluid codes (e.g. SOLPS), including underestimating outer target temperatures\cite{Chankin2009a}, radial electric field in the SOL\cite{Chankin2007,Chankin2009,Chankin2009a}, parallel ion SOL flows at the low field side\cite{Asakura2007,Chankin2009,Groth2009}, and impurity radiation\cite{Canik2015}. It was hypothesized by Chankin et al.\cite{Chankin2009} that these discrepancies stem from the ad-hoc treatment of kinetic effects in fluid codes.

To determine the importance of kinetic effects in the SOL, simulations were undertaken using the XGCa code\cite{Hager2016}, a total-$f$, gyrokinetic code which self-consistently calculates the axisymmetric electrostatic potential and plasma dynamics, and includes modules for Monte Carlo neutral transport. General features of the simulation results are investigated for kinetic effects, but also in the future comparisons will be made to the fluid SOL transport code SOLPS\cite{Rozhansky2009}. We note here that the present study is without turbulence.  Turbulence may alter the results presented here.

Here we focus on the question of parallel pressure balance in a low-density, low-power DIII-D H-mode plasma. A significant deviation from standard total pressure balance ($p_e + p_i + m_i n_i V_{i,\parallel}^2 = const$) occurs in the SOL of these XGCa simulations, and here we explore whether the addition of main ion viscosity and neutral source terms can account for this deviation.

The rest of the paper is organized as follows: Section \ref{sec:xgca} briefly describes the XGCa code, Section \ref{sec:sim_setup} details some of the simulation setup, Section \ref{sec:pres_bal} discusses details of the pressure balance parallel to magnetic field lines, including main ion viscosity and neutral source terms, and Section \ref{sec:conclusion} wraps up with conclusions and plans for future work.

\section{XGCa}\label{sec:xgca}
XGCa is a total-\emph{f}, gyrokinetic neoclassical particle-in-cell (PIC) code\cite{Hager2016,Ku2009,Ku2016}. The ions are pushed according to a gyrokinetic formalism, and the electrons are drift-kinetic. XGCa is very similar to the more full featured, gyrokinetic turbulence version XGC1\cite{Ku2009,Chang2009a,Ku2016}, the main difference being that XGCa solves only for the axisymmetric electric potential (i.e. no turbulence, hence the "neoclassical" descriptor). An important feature of XGCa is that the poloidally varying electric potential is calculated by solving a gyrokinetic Poisson equation, so that the resulting electric field is self-consistent with the gyrating kinetic particles. XGCa also uses a realistic magnetic geometry, created directly from experiment magnetic reconstructions (normally from EFIT EQDSK files), including X-points and material walls. Particle drifts (magnetic and $E \times B$) are included on the particle motion. Kinetic effects of neutrals from charge-exchange and ionization are included, along with consistent neutral rates set by global recycling. The Debye sheath region isn't resolved, but rather a modified logical sheath boundary condition\cite{Churchill2016, Parker1993} is used to impose ambipolar flux to the material walls. There are two main differences between XGCa and the previous neoclassical version XGC0.  XGC0 calculates only flux-averaged potential and includes the gyro-averaging effect in a simplified manner.

\section{XGCa Simulation Setup}\label{sec:sim_setup}
The results presented in this paper are from an XGCa simulation of a low-power H-mode discharge of the DIII-D tokamak\cite{Luxon2005}, shot 153820 at time 3000 ms. The simulation parameters, including experimental inputs of ion and electron density and temperature, can be found in Ref. \cite{Churchill2016}. This discharge was chosen for its lower density, so that the SOL collisionality would be low, where kinetic effects would be expected to be more significant (so called "sheath-limited" regime).

\section{Pressure Balance}\label{sec:pres_bal}
Pressure balance in the SOL ($p_e + p_i + m_i n_i V_i^2 = const$) is often used to derive expected target plasma profiles, for example using two-point modeling\cite{Stangeby2000}. For small midplane ion flows, and target flows satisfying the Bohm criterion ($V_{i,\parallel} = c_s$), the midplane static pressure ($p_e + p_i$) should be 2x the divertor static pressure\cite{Stangeby2000,Paradela2016}. 
SOL measurements of upstream $n_e$, $T_e$, and $T_i$, with downstream $n_e$ and $T_e$ done in DIII-D attached ELMing H-mode plasmas showed\cite{Petrie1992} seem to agree with this. However, assumptions of low midplane flows and $T_i \approx T_e$ in the divertor had to be made, since main ion measurements are notoriously difficult to measure in the SOL generally.

\begin{figure}%[htbp]
\centering
\includegraphics[width=0.45\textwidth]{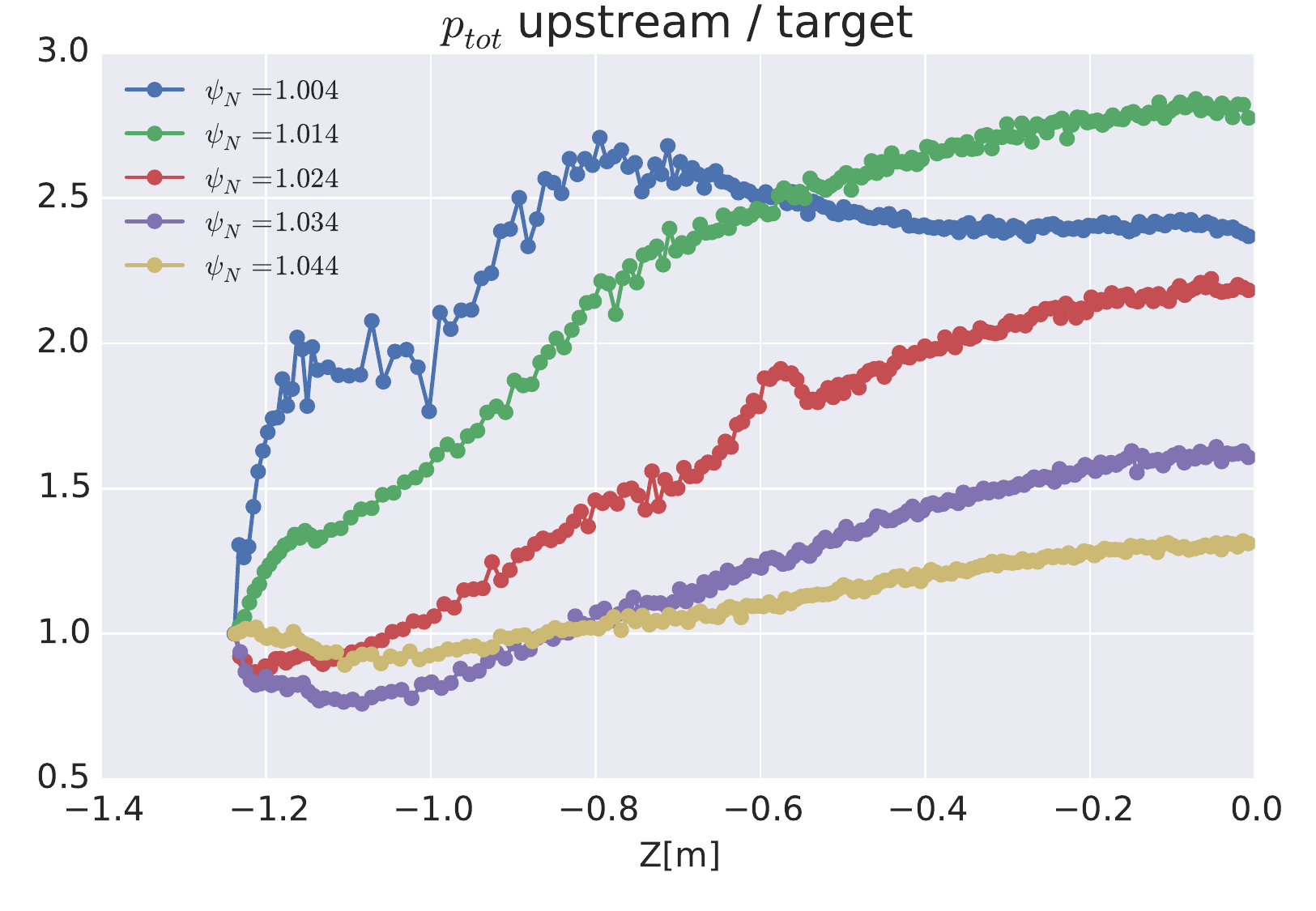}
\caption{The ratio (upstream over target) of total pressure $(p_{tot} = p_e + p_i + m_i n_i V_{i\parallel}^2)$ between the LFS divertor target and upstream midplane, plotted versus height (Z). The X-point is located at Z=-1.14.}
\label{fig:presbal_simple}
\end{figure}

Previously presented XGCa results showed large variations of ion density and pressure along field lines in the SOL, and significant parallel ion flows (Mach number 0.5) throughout the SOL\cite{Churchill2016}. A natural check was to ensure pressure balance was satisfied. As a simple test, the total fluid pressure ($p_{tot} = p_e + p_i + m_i n_i V_{i\parallel}^2$) in the SOL at the low-field side (LFS) from this XGCa simulation is plotted in Figure \ref{fig:presbal_simple}, normalized to the LFS target total pressure. A large imbalance is seen, ranging from near balanced in the farthest out flux surfaces to factors of 2.5 for flux surfaces in the near-SOL.

Here we consider adding the momentum drag from neutrals and the main ion viscosity into the parallel momentum equation, to determine if this will balance the momentum equation between the target and upstream. As will be seen, these simple fluid terms cannot account for the pressure imbalance from the kinetic simulation, requiring future work on a more complete fluid momentum balance equation.

\subsection{Parallel Momentum Equation}
A total parallel momentum equation\cite{Churchill2014,Helander2005} can be formed using the procedure and assumptions outlined in Ref. \cite{Churchill2014}, arriving at:

%for arbitrary plasma species $j$ is:
%\begin{equation}
%m_{j}n_{j}\left[\frac{\partial\mathbf{V}_{j}}{\partial %t}+\mathbf{V}_{j}\cdot\nabla\mathbf{V}_{j}\right]=-\nabla %p_{j}-\nabla\cdot\overset\leftrightarrow{\pi}_{j}+Z_{j}en_{j}\left(\mathbf{E}+\mathbf{V}_{j}\times\%mathbf{B}\right)+\mathbf{R}_{j}
%\end{equation}

%\begin{equation}
%\begin{array}{cc}
%\begin{multline}
%\mathbf{b} \cdot \left[ m_{i}n_{i}\mathbf{V}_{i} \cdot \nabla \mathbf{V}_{i} + %\nabla(p_{e}+p_{i})+\nabla \cdot \mathbf{\pi}_{i} \right]  \\ +  
% m_i V_{i,\parallel} n_n \left(\nu_{ion} + \nu_{cx} \right) = 0
%\end{multline}
%\end{align}
%\end{equation}
\begin{eqnarray*}
\fl \mathbf{b} \cdot [ m_{i}n_{i}\mathbf{V}_{i} \cdot \nabla \mathbf{V}_{i}  & + \nabla(p_e+p_i)+\nabla \cdot \mathbf{\pi}_i ] + &  \\ 
& m_i V_{i,\parallel} n_n \left(\nu_{ion} + \nu_{cx} \right) & = 0
\label{eq:eq3}
\end{eqnarray*}

where $\nu_{ion} = n_e \langle \sigma_{ion} v \rangle$ and $\nu_{cx} = n_i \langle \sigma_{cx} v \rangle$ are the ionization and charge exchange rates. Elastic collisions were turned off, as they should only become important for cold, detached divertors\cite{Kanzleiter2000}. We further simplify the ion inertia term as $\mathbf{b} \cdot \mathbf{V}_i \cdot \nabla \mathbf{V}_i = \mathbf{b} \cdot \nabla(V_i^2/2) - \mathbf{V}_i \times \nabla \times \mathbf{V}_i \approx \mathbf{b} \cdot \nabla (V_i^2/2)$, noting that $\mathbf{V}_i \times \nabla \times \mathbf{V}_i$ can contain important terms. We also simplify the viscosity using only the parallel viscosity (neglecting perpendicular gyroviscosity) in the Chew-Golberger-Low form\cite{Helander2005,Shaing1993}, $\mathbf{b} \cdot \nabla \cdot \pi_{i\parallel} = \frac{2}{3}\mathbf{b} \cdot \nabla (p_{i\parallel} - p_{i\perp})  - (p_{i\parallel} - p_{i\perp}) \mathbf{b} \cdot \nabla \ln B$. Substituting these simplifications, and integrating over the parallel direction from $\ell_\parallel = 0$ at the divertor target to $x$, an upstream value, we arrive at the final pressure balance in Equation \ref{eq:presbal_all}.:

%\begin{multline}
%\mathbf{b} \cdot \nabla \left( p_e + p_i + \frac{1}{2} m_i n_i V_i^2 + %\frac{2}{3}(p_{i\parallel} - p_{i\perp}) \right) \\ +
%m_i V_{i\parallel} n_n \left(\nu_{ion} + \nu_{cx}\right) \\ - 
%(p_{i\parallel} - p_{i\perp}) \mathbf{b} \cdot \nabla \ln B = 0
%\end{multline}
%\begin{equation}
%\centering
%\mathbf{b} \cdot \nabla \left( p_{tot} + \frac{2}{3}(p_{i\parallel} - p_{i\perp}) \right)  +
%m_i V_{i\parallel} n_n \left(\nu_{ion} + \nu_{cx}\right) - 
%(p_{i\parallel} - p_{i\perp}) \mathbf{b} \cdot \nabla \ln B = 0
%\end{equation}

%\begin{strip}
\begin{equation}
\centering
\frac{p_{tot} |_{\ell_\parallel = x}}{p_{tot} |_{\ell_\parallel = 0}} + F_{visc} + F_{neu} = 1
%\frac{(p_{tot} + \frac{2}{3}(p_{i\parallel} - p_{i\perp}))|_{\ell_\parallel = x} + \int_{0}^{x} %d\ell_\parallel \, [ m_i V_{i\parallel} n_n (\nu_{ion} + \nu_{cx}) - (p_{i\parallel} - %p_{i\perp}) \mathbf{b} \cdot \nabla \ln B }
 %{(p_{tot} + \frac{2}{3}(p_{i\parallel} - p_{i\perp}) )|_{\ell_\parallel = 0}} = 1
\label{eq:presbal_all}
\end{equation}
%\end{strip}

where we have written:
\begin{eqnarray}
p_{tot} = p_e + p_i + \frac{1}{2} m_i n_i V_i^2 \\
F_{visc} = \frac{\frac{2}{3}(p_{i\parallel} - p_{i\perp})|^{\ell_\parallel = x}_{\ell_\parallel = 0} - \int_{0}^{x} d\ell_\parallel \, [(p_{i\parallel} - p_{i\perp}) \mathbf{b} \cdot \nabla \ln B}
{p_{tot}|_{\ell_\parallel = 0}} \\
F_{neu} = \frac{\int_{0}^{x} d\ell_\parallel m_i V_{i\parallel} n_n (\nu_{ion} + \nu_{cx})}
{p_{tot}|_{\ell_\parallel = 0}}
\label{eq:presbal_2}
\end{eqnarray}

\subsection{Ion temperature anisotropy}
\begin{figure}[!htb]
%\begin{wrapfigure}[11]{R}{0.5\textwidth}
\centering
\includegraphics[width=\columnwidth]{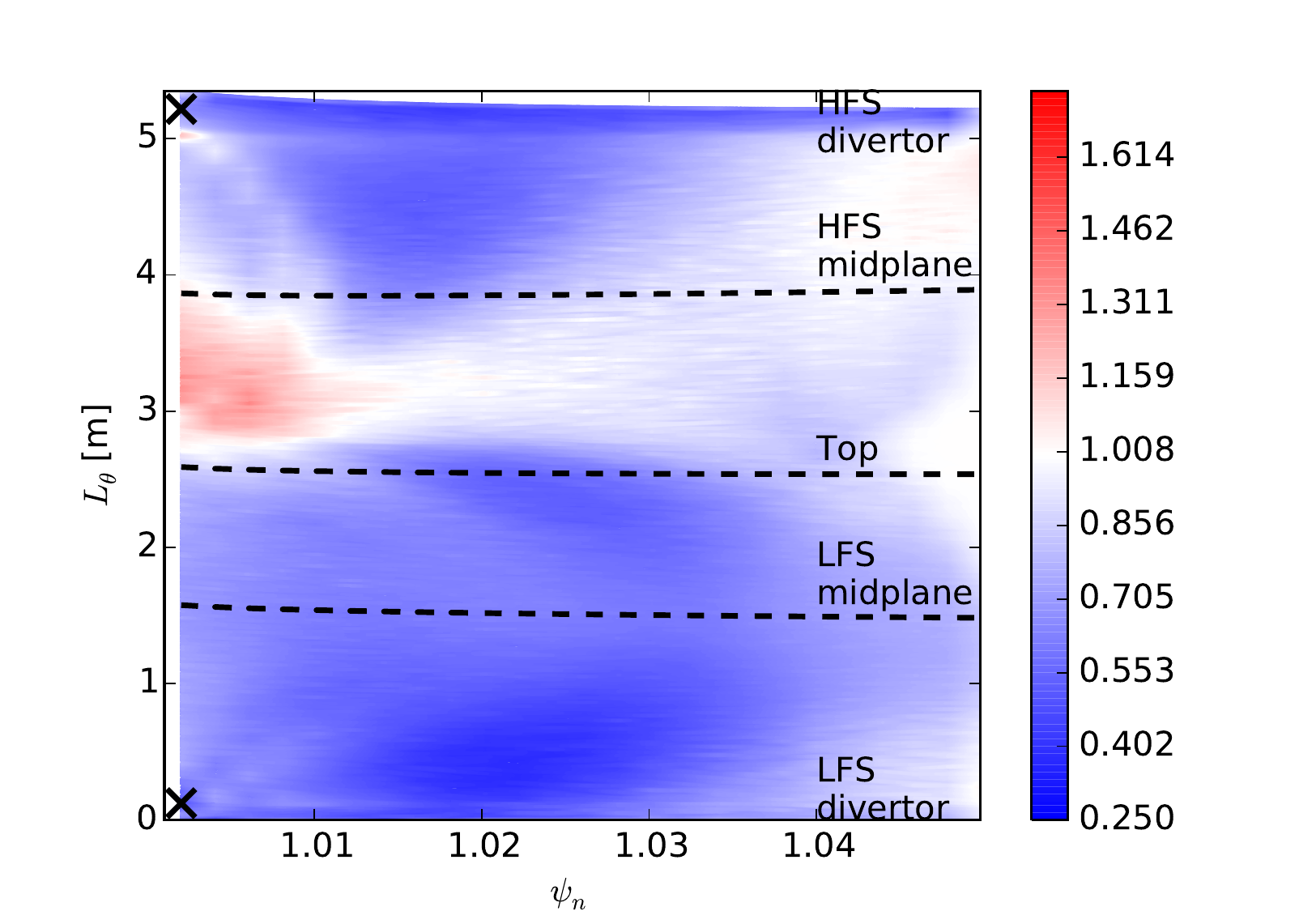}
\caption{Ion temperature anisotropy ratio, $T_{i\parallel} / T_{i\perp}$. X-point locations are marked with the black X's.}
\label{fig:ti_anisotropy}
\end{figure}

Ion parallel and perpendicular pressure anisotropy is the drive in the parallel viscosity term, which is not normally considered in fluid codes since strong collisionality would generally reduce anisotropies in confined plasmas. The kinetic XGCa simulation results show that there is a strong main ion temperature anisotropy in the SOL, beginning just inside the separatrix in the pedestal region, and increasing in the SOL. The plot of SOL $T_{i\parallel} / T_{i\perp}$ in Figure \ref{fig:ti_anisotropy} shows the ion temperature (or more precisely the average kinetic energy) anisotropy ratio reaches levels of 0.25, but also has regions where $T_{i\parallel} > T_{i\perp}$, with the ratio as high as 1.35 at the top of the machine. The LFS region ion temperature anisotropy can be explained by a significant trapped ion fraction, which leads to a predominantly higher perpendicular energy. The region of $T_{i\parallel} > T_{i\perp}$ near the top is also due to kinetic effects with the higher parallel-energy ions flowing more freely to the top avoiding trapping at LFS. The low end anisotropy ratio is near the simple theoretical limit of ion viscosity\cite{Stangeby2000} ($\pi_i > -p_i$), which translates to $p_{i\parallel} / p_{i\perp} > 2/5$.

The ion viscosity in this case works to decrease the total pressure (it's identically negative), and does contribute a non-negligible amount to the pressure balance. It dominantly affects flux surfaces in the middle of the SOL $\psi_n$ range.

Note, however, that it does not fix the total pressure balance discrepancy in Figure \ref{fig:presbal_simple}.

\begin{figure}[!htb]
\centering
\includegraphics[width=\columnwidth]{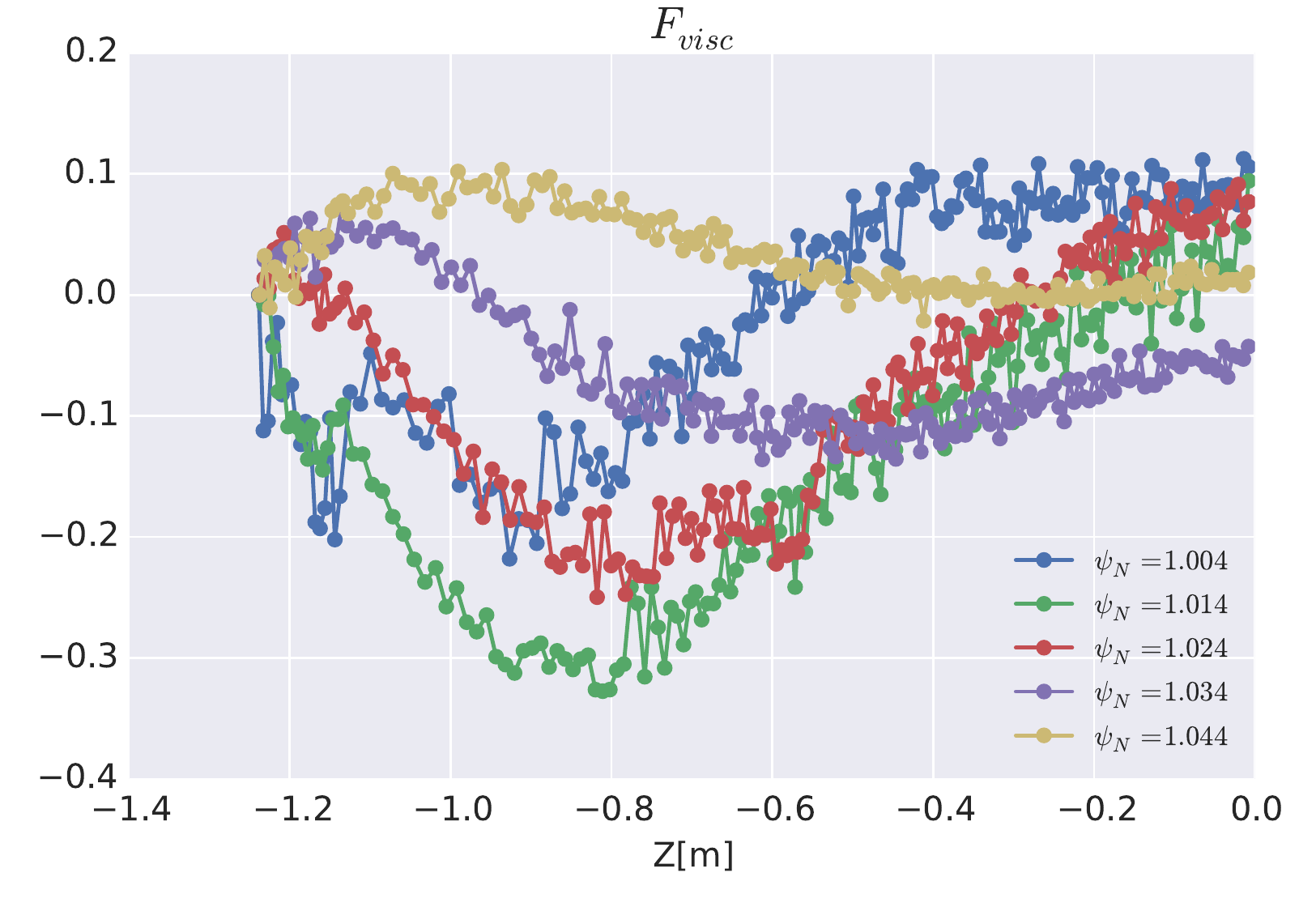}
\caption{Main ion viscosity contribution ($F_{visc}$) to the total pressure balance, between the LFS divertor target and midplane, plotted versus height (Z). The X-point is located at Z=-1.14.}
\label{fig:pii_over_pres}
\end{figure}

\subsection{Neutral source}
\begin{figure}[!htb]
\centering
\includegraphics[width=\columnwidth]{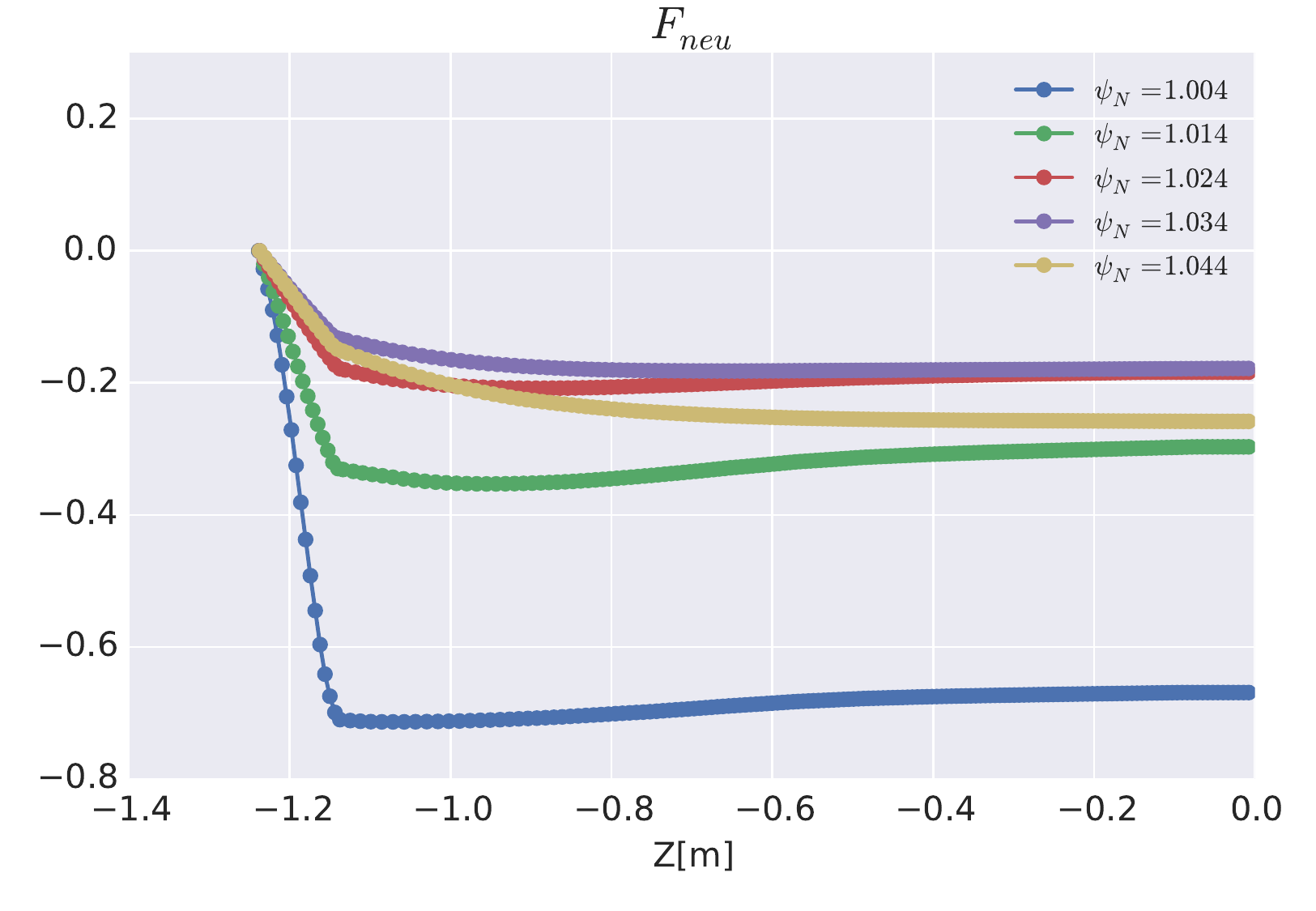}
\caption{Neutral source contribution ($F_{neu}$) to total pressure balance, between the LFS divertor target and midplane, plotted versus height (Z). The X-point is located at Z=-1.14.}
\label{fig:source_over_pres}
\end{figure}

Having investigated the main ion viscosity, we now turn our attention to the neutral momentum term, which acts as a drag. This is due to dominantly colder neutrals joining the ion population through charge-exchange or ionization. Shown in Figure \ref{fig:source_over_pres} is the neutral term contribution to the total pressure balance,  $\int m_i V_i n_n (\nu_{ion} + \nu_{cx}) / (\mathbf{b} \cdot \nabla (p_e + p_i + m_i n_i V_{i\parallel}^2 + 2/3(p_{i\parallel} - p_{i\perp}) )|_{target}$. This term can be substantial, especially for near-SOL flux surfaces. It also, however, does not fix the total pressure balance discrepancy in Figure \ref{fig:presbal_simple}.

\subsection{Pressure balance updated}
\begin{figure}%[htbp]
\centering
\includegraphics[width=\columnwidth]{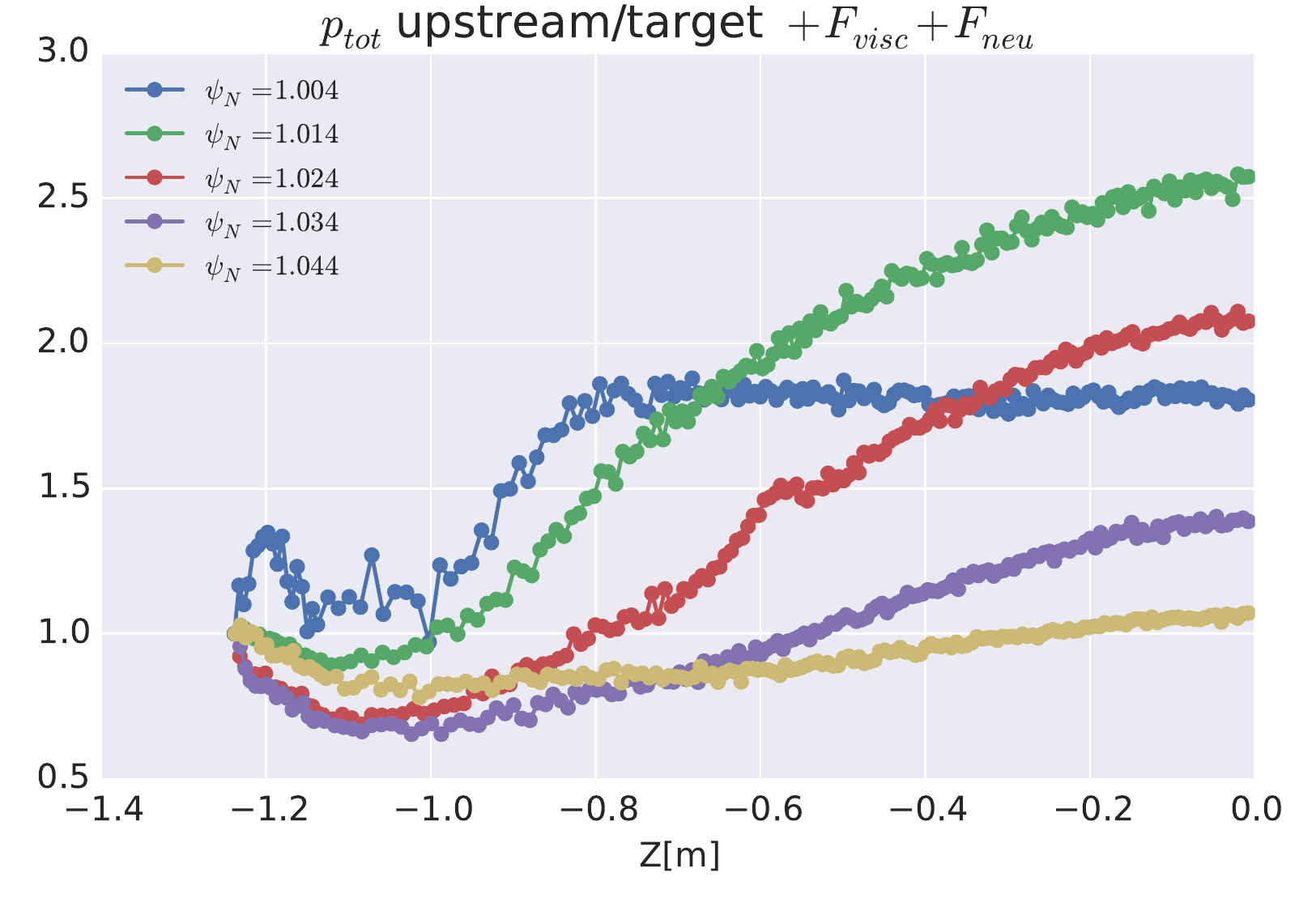}
\caption{The pressure balance using Equation \ref{eq:presbal_all}, showing that adding main ion viscosity and neutral sources cannot account for the total pressure imbalance. The X-point is located at Z=-1.14.}
\label{fig:presbal_all}
\end{figure}

Using the contributions from the main ion viscosity and neutral source, we update the plot of Figure \ref{fig:presbal_simple}, to plot the more complete pressure balance equation, Equation \ref{eq:presbal_all} in Figure \ref{fig:presbal_all}. As seen, there is a significant deviation from 1 for flux surfaces in the near- to mid-SOL, but only for regions above $Z>-1.0$ m (recall the X-point is at $Z=-1.14$ m). For the flux surfaces further in the far-SOL, where the collisionality is higher, the agreement is fairly good, deviating on the level of 10\% between the target and midplane.

The pressure imbalance is dominated by the variation in the ion temperature, and so is likely due to not including kinetic effects from high energy ion contributions not included in this fluid picture. It is possible that including the in-surface flows in the inertia term ($\mathbf{V} \times \nabla \times \mathbf{V}$) can also be important, and needs to be explored.

\section{Conclusions and Future Work}\label{sec:conclusion}
The XGC codes are useful for probing and predicting kinetic effects in the edge region, including the scrape-off layer, as they include many of the interconnected physics necessary for realistic modelling. An XGCa simulation of a DIII-D H-mode showed several novel features. The ion density and temperature are larger at the LFS, indicating effects from fat ion orbits from the confined pedestal region. The parallel ion Mach number at the LFS midplane reaches experimental levels ($M_i \sim 0.5$), and shows a poloidal variation consistent with the parallel ion flows being dominated by Pfirsch-Schl{\"u}tter flows (recall XGCa includes drifts), with stagnation points near the X-point at both the LFS and HFS and flows directed towards the divertor below the X-point. The normalized sheath potential at the divertor plates is higher than standard textbook assumptions, along with subsonic ions at the sheath edge, which both implicate kinetic effects in the establishment of the sheath potential in this discharge.

Pressure balance in the SOL of these simulations was explored, showing a strong deviation when using total pressure only. Significant ion temperature anisotropies are present in the pedestal and SOL of this simulation, indicating main ion viscosity could play a significant role in the pressure balance. Main ion viscosity and neutral source terms were included in the momentum equation, and, while significant in the SOL, they could not account for the entire fluid pressure imbalance seen in XGCa , especially for flux surfaces in the near- and mid-SOL.

The present physics results are without turbulence or impurity particles, leaving them for future work. Turbulence and impurities could modify the results presented here.

\section{Acknowledgements}
This work supported by the U.S. Department of Energy under DE-AC02-09CH11466, DE-AC05-00OR22725, and DE-FC02-04ER54698.  Awards of computer time was provided by the Innovative and Novel Computational Impact on Theory and Experiment (INCITE) program. This research used resources of the Argonne Leadership Computing Facility, which is a DOE Office of Science User Facility supported under contract DE-AC02-06CH11357. This research also used resources of the National Energy Research Scientific Computing Center, a DOE Office of Science User Facility supported by the Office of Science of the U.S. Department of Energy under Contract No. DE-AC02-05CH11231. DIII-D data shown in this paper can be obtained in digital format by following the links at https://fusion.gat.com/global/D3D\_DMP

\bibliography{iaea2016}
%\printbibliography %used with biblatex, incompatible with natbib

\end{document}